\documentclass[aps,prb,reprint,preprintnumbers,superscriptaddress,amsmath,amssymb,bibnotes,longbibliography]{revtex4-2}
\pdfoutput=1
\usepackage{graphicx}
\usepackage{bm}

\graphicspath{{figures/}{../}}
\usepackage{color}
\usepackage[colorlinks,bookmarks=false,citecolor=darkblue,linkcolor=red,urlcolor=blue]{hyperref}

\definecolor{darkred}{rgb}{0.7,0.0,0.0}
\definecolor{darkblue}{rgb}{0,0.02,0.45}

\definecolor{darkgreen}{rgb}{0.02,0.45,0.0}

\begin{document}

\title{Magnetic versus nonmagnetic polymorphs of RuBr$_3$ under pressure}

\author{Bin Shen}
\altaffiliation{These authors contributed equally}
\affiliation{Experimental Physics VI, Center for Electronic Correlations and Magnetism, University of Augsburg, 86159 Augsburg, Germany}

\author{Victoria A. Ginga}
\altaffiliation{These authors contributed equally}
\affiliation{Felix Bloch Institute for Solid-State Physics, University of Leipzig, 04103 Leipzig, Germany}

\author{Angel M. Ar\'evalo-L\'opez}
\affiliation{UMR-8181-UCCS-Unit\'e de Catalyse et Chimie du Solide-Univ. Lille, CNRS, Centrale Lille, ENSCL, Univ. Artois, F-59000 Lille, France}

\author{Gaston Garbarino}
\affiliation{European Synchrotron Radiation Facility, 38043 Grenoble, France}

\author{Ece Uykur}
\affiliation{Helmholtz Zentrum Dresden Rossendorf, Inst Ion Beam Phys \& Mat Res, D-01328 Dresden, Germany}

\author{Marcos Gon{\c c}alves-Faria}
\affiliation{Helmholtz Zentrum Dresden Rossendorf, Inst Ion Beam Phys \& Mat Res, D-01328 Dresden, Germany}

\author{Prashanta K. Mukharjee}
\affiliation{Experimental Physics VI, Center for Electronic Correlations and Magnetism, University of Augsburg, 86159 Augsburg, Germany}


\author{Philipp Gegenwart}
\affiliation{Experimental Physics VI, Center for Electronic Correlations and Magnetism, University of Augsburg, 86159 Augsburg, Germany}

\author{Alexander A. Tsirlin}
\email{altsirlin@gmail.com}
\affiliation{Felix Bloch Institute for Solid-State Physics, University of Leipzig, 04103 Leipzig, Germany}

\begin{abstract}
Pressure evolution of the crystal structure and magnetism of the honeycomb $\alpha$-RuBr$_3$ is studied using high-pressure x-ray diffraction, magnetometry, and density-functional band-structure calculations. Hydrostatic compression transforms antiferromagnetic $\alpha$-RuBr$_3$ ($R\bar 3$) into paramagnetic \mbox{$\alpha'$-RuBr$_3$} ($P\bar 1$) where short Ru--Ru bonds cause magnetism collapse above 1.3\,GPa at 0\,K and 2.5\,GPa at 295\,K. Below this critical pressure, the N\'eel temperature of $\alpha$-RuBr$_3$ increases with the slope of 1.8\,K/GPa. Pressure tunes $\alpha$-RuBr$_3$ away from the Kitaev limit, whereas increased third-neighbor in-plane coupling and interlayer coupling lead to a further stabilization of the collinear zigzag state. Both $\alpha$- and $\alpha'$-RuBr$_3$ are metastable at ambient pressure, but their transformation into the thermodynamically stable $\beta$-polymorph is kinetically hindered at room temperature. 
\end{abstract}

\maketitle

\section{Introduction}
Experimental realization of the Kitaev model and its intriguing spin-liquid physics~\cite{kitaev2006} requires honeycomb magnets with the $d^5$ or $d^7$ transition-metal ions~\cite{jackeli2009,liu2018}. The choice of chemical compounds satisfying these criteria appears to be quite limited, especially in the case of Ru$^{3+}$ ($4d^5$) that has been known to form only one honeycomb magnet, the widely studied \mbox{$\alpha$-RuCl$_3$}~\cite{winter2017,takagi2019}. Other ruthenium trihalides exist too, but they adopt chain-like structures and show mundane temperature-independent paramagnetic behavior~\cite{schnering1966}. A similar chain-like structure reported for the chloride is commonly identified as $\beta$-RuCl$_3$ in contrast to the $\alpha$-polymorph with the honeycomb layers~\cite{fletcher1963,fletcher1967,kobayashi1992}.

Relative stability of the $\alpha$- and $\beta$-polymorphs of RuCl$_3$ is controlled by temperature. Whereas $\beta$-RuCl$_3$ is synthesized at $600-650$\,K, increasing the synthesis temperature above 700\,K stabilizes the magnetic $\alpha$-polymorph~\cite{hyde1965}. A somewhat similar transformation is also possible in the bromide. Its chain-like paramagnetic form, $\beta$-RuBr$_3$, is the only polymorph that can be synthesized at ambient pressure~\cite{hillebrecht2004}. However, the high-pressure high-temperature treatment of $\beta$-RuBr$_3$ leads to the honeycomb structure of $\alpha$-RuBr$_3$~\cite{imai2022,prots2023} that shares many similarities with $\alpha$-RuCl$_3$, including its local magnetism of Ru$^{3+}$ and collinear zigzag magnetic order at low temperatures~\cite{imai2022,choi2022,weinhold2024}. External pressure and temperature can thus control the formation of magnetic vs. nonmagnetic polymorphs of the Ru$^{3+}$ trihalides.

\begin{figure*}
\includegraphics{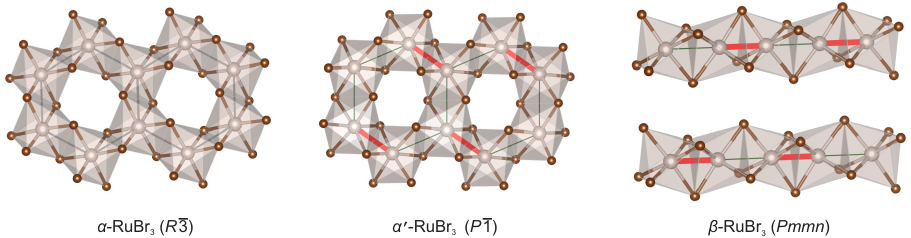}
\caption{\label{fig:structure}
Crystal structures of the RuBr$_3$ polymorphs. The red lines show the Ru--Ru dimers.
}
\end{figure*}

High-pressure treatment renders RuBr$_3$ magnetic. This evolution is remarkably different from $\alpha$-RuCl$_3$ and honeycomb iridates that typically become nonmagnetic upon application of pressure~\cite{bastien2018,shen2021,shen2022}, thus limiting the use of pressure as a tuning parameter. The magnetism collapse in all these Kitaev materials is caused by the formation of short metal-metal bonds that break regular honeycombs into nonmagnetic dimers~\cite{hermann2018,biesner2018,takayama2019,veiga2019}. The same mechanism is in fact responsible for the temperature-independent paramagnetic behavior of $\beta$-RuBr$_3$ where metal-metal bonds (dimers) are formed within the ruthenium chains~\cite{brodersen1968,merlino2004,hillebrecht2004,prots2023}, see Fig.~\ref{fig:structure}. This collapsed magnetic state is suppressed when $\beta$-RuBr$_3$ transforms into the $\alpha$-polymorph upon application of pressure. It raises an interesting question whether $\alpha$-RuBr$_3$ may be more robust against pressure-induced magnetism collapse than $\alpha$-RuCl$_3$, thus offering a broader pressure window for tuning Kitaev magnetism in the honeycomb planes. The larger unit-cell volume of $\alpha$-RuBr$_3$ compared to $\alpha$-RuCl$_3$ should also facilitate the stability of this compound against the pressure-induced structural dimerization.

In the following, we investigate $\alpha$-RuBr$_3$ under hydrostatic pressure and report its structural evolution as well as magnetic behavior. We show that $\alpha$-RuBr$_3$ follows the same scenario of pressure-induced structural dimerization as the chloride, but with the magnetism collapse happening at higher pressures. This similar behavior contrasts with the different pressure evolution of the N\'eel temperature $T_N$ in the chloride and bromide. We also assess thermodynamic stability of the RuBr$_3$ polymorphs and reveal the metastable nature of $\alpha$-RuBr$_3$ at ambient pressure.

\section{Methods}
Polycrystalline samples of $\alpha$-RuBr$_3$ were prepared by annealing the commercial $\beta$-RuBr$_3$ powder (Alfa Aesar) at 6\,GPa and 900\,$^{\circ}$C in a Walker-type multianvil press followed by cooling the sample to room temperature before releasing the pressure. The recovered powder contained the $R\bar 3$ phase ($\alpha$-RuBr$_3$) with 1.6\,wt.\% of the RuO$_2$ impurity according to the Rietveld refinement of x-ray diffraction (XRD) data. No traces of the $\beta$-polymorph were detected.

High-pressure XRD data were collected at room temperature at the ID27 beamline of the European Synchrotron Radiation Facility (ESRF, Grenoble, France) using the wavelength of 0.3738\,\r A and EIGER2 X CdTe 9M detector. Powder samples of $\alpha$-RuBr$_3$ and $\beta$-RuBr$_3$, respectively, were placed into a stainless-steel gasket mounted inside a diamond anvil cell filled with helium gas as pressure-transmitting medium. Pressure was measured using the fluorescence line of a ruby sphere placed into the cell next to the sample. Two-dimensional images were integrated using the \texttt{Dioptas} software~\cite{dioptas}. \texttt{Jana2006}~\cite{jana2006} was used for structure refinement. The high-pressure XRD data were collected up to 17\,GPa for $\alpha$-RuBr$_3$ and up to 12\,GPa for $\beta$-RuBr$_3$.

Magnetization under pressure was measured similar to Ref.~\cite{shen2021}. Pressure was calibrated by measuring the superconducting transition of Pb. Daphne oil 7373 was used as pressure-transmitting medium. The data were collected in two runs performed on different portions of the same $\alpha$-RuBr$_3$ sample. The 1.8\,mm anvil-culet and the gasket with the sample space diameter of 0.9\,mm were used.

Density-functional (DFT) band-structure calculations were performed in the \texttt{VASP} code~\cite{vasp1,vasp2} using the Perdew-Burke-Ernzerhof (PBE) exchange-correlation potential~\cite{pbe96} with Grimme's D3 dispersion correction~\cite{grimme2010} for weak van der Waals bonding, which is expected in the $\alpha$- and $\beta$-polymorphs between the layers and chains, respectively. Crystal structures of different polymorphs were optimized at several constant volumes to obtain the equation of state and calculate enthalpies as a function of pressure. Additionally, \texttt{FPLO}~\cite{fplo} was used to calculate PBE band structures of $\alpha$-RuBr$_3$ at different pressures using experimental lattice parameters. Atomic positions were optimized at each pressure prior to constructing the tight-binding models via Wannier projections~\cite{koepernik2023} and estimating exchange couplings using the superexchange model developed in Refs.~\cite{rau2014,winter2016}. First Brillouin zone was sampled by a fine mesh with up to 700 $k$-points in the symmetry-irreducible part.

\section{Results}
\subsection{Crystal structure}

\begin{figure*}
\includegraphics{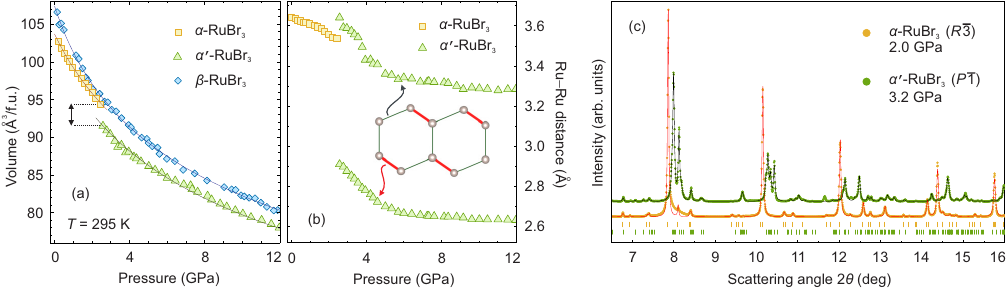}
\caption{\label{fig:volume}
(a) Pressure dependence of the unit-cell volume measured by XRD. The solid lines are the fits with Eq.~\eqref{eq:eos2}. The dotted lines hightlight the volume drop upon the $\alpha-\alpha'$ transition. (b) Pressure dependence of the Ru--Ru distances. The two longer distances in the $\alpha'$-polymorph have been averaged for clarity. (c) XRD patterns of $\alpha$- and $\alpha'$-RuBr$_3$ (offset for clarity). The lines show the Le Bail fits to the data.
}
\end{figure*}

Our $\alpha$-RuBr$_3$ sample shows the $R\bar 3$ crystal structure at ambient pressure in agreement with the previous report~\cite{imai2022}. This rhombohedral structure remains stable up to 2.5\,GPa at room temperature. At higher pressures, an abrupt change in the XRD patterns (Fig.~\ref{fig:volume}c) indicates a phase transition toward the high-pressure polymorph with lower symmetry. This $\alpha'$-RuBr$_3$ polymorph is triclinic ($P\bar 1$), similar to the high-pressure dimerized phase of RuCl$_3$~\cite{bastien2018}. No further pressure-induced transformations were observed up to 17\,GPa, the highest pressure of our experiment~\cite{supplement}. The $\alpha-\alpha'$ transformation is fully reversible. Upon releasing pressure, the $\alpha$-polymorph was recovered, albeit with the increased diffuse scattering indicative of the higher number of stacking faults that develop in the weakly bonded layered structure upon pressure cycling~\cite{supplement}.

The pressure-induced structural phase transition is accompanied by the volume drop of about 3\% (Fig.~\ref{fig:volume}a), which is comparable to the 2.3\% volume drop in $\beta$-Li$_2$IrO$_3$~\cite{veiga2019} and 1.9\% in $\alpha$-Li$_2$IrO$_3$~\cite{hermann2018} upon their pressure-induced phase transitions. The Ru--Ru distances extracted from the structure refinements show that RuBr$_3$ also undergoes a structural dimerization, resulting in the magnetism collapse (see Sec.~\ref{sec:magnetic}). Three equivalent Ru--Ru distances of about 3.6\,\r A in $\alpha$-RuBr$_3$ split into one short and two long, almost equal distances in the high-pressure phase (Fig.~\ref{fig:volume}b). The shorter distance is about 2.9\,\r A right above the transition and still exceeds the Ru--Ru distance of $2.71-2.73$\,\r A in $\beta$-RuBr$_3$~\cite{hillebrecht2004,prots2023}, but this dimer distance in $\alpha'$-RuBr$_3$ rapidly shrinks and goes below 2.7\,\r A above 5\,GPa. The rapid reduction in the Ru--Ru distances between 2.5 and 5\,GPa may be caused by the large size of bromine that increases the average Ru--Ru distance in $\alpha$-RuBr$_3$ compared to the chloride and makes it more difficult to form the dimers. Higher pressure is thus required to complete the dimer formation.

A comparative pressure-dependent study of $\beta$-RuBr$_3$ showed that this polymorph does not transform into the layered ($\alpha$- or $\alpha'$-) structure up to at least 12\,GPa at room temperature. It undergoes a steady compression~\cite{supplement} with a similar pressure dependence as in $\alpha'$-RuBr$_3$, albeit with the larger unit-cell volume (Fig.~\ref{fig:volume}a). Below we show that the $\beta-\alpha'$ transition should be thermodynamically favored above 5.5\,GPa, and indeed a transformation into the layered polymorph occurs upon the high-pressure high-temperature treatment. The persistence of $\beta$-RuBr$_3$ up to much higher pressures at room temperature indicates that this transformation is kinetically hindered.

\subsection{Thermodynamics}

\begin{figure}
\includegraphics{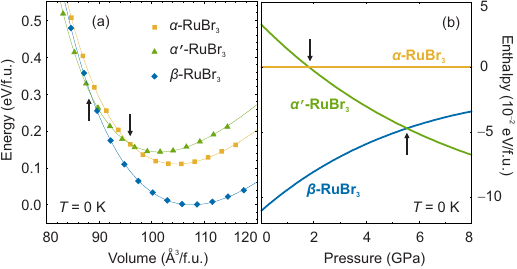}
\caption{\label{fig:energies}
(a) Total energies of the RuBr$_3$ polymorphs as a function of volume given relative to the energy minimum of the most stable polymorph, $\beta$-RuBr$_3$. The lines are the fits with Eq.~\eqref{eq:eos}. (b) Pressure dependence of the enthalpies given relative to $\alpha$-RuBr$_3$. The arrows show the $\alpha-\alpha'$ and $\beta-\alpha'$ transitions
}
\end{figure}

To assess thermodynamic stability of the different polymorphs, we calculated their total energies at several fixed volumes (Fig.~\ref{fig:energies}a) and fitted these energies to the Murnaghan equation of state, 
\begin{align}
	E(V)=E_0+B_0V_0 &\left[\frac{1}{B_0'(B_0'-1)}\right. \left(\frac{V}{V_0}\right)^{1-B_0'}+ \notag\smallskip \\
	&\left.+\frac{1}{B_0'}\frac{V}{V_0}-\frac{1}{B_0'-1}\right]
\label{eq:eos}\end{align}
where $B_0$ is the bulk modulus at ambient pressure, $B_0'$ shows linear pressure dependence of the bulk modulus, while $E_0$ and $V_0$ are the equilibrium energy and volume, respectively. These parameters listed in Table~\ref{tab:eos} show a good agreement with the experimental values obtained by fitting the $V(P)$ curves to 
\begin{equation}
 V(P)=V_0\left(\frac{B_0'}{B_0}P+1\right)^{-1/B_0'}
\label{eq:eos2}\end{equation}
where the $V_0$ parameter was kept fixed for each of the $\alpha$- and $\alpha'$- polymorphs because of the limited pressure window available for these structures.

\begin{table}
\caption{\label{tab:eos}
Parameters of the equation of state derived from the experimental pressure-dependent unit-cell volume, $V(P)$, and from total energies $E(V)$ calculated by DFT. The parameters $E_0$ and $V_0$ stand for the equilibrium energy (calculated with respect to the most stable polymorph, $\beta$-RuBr$_3$) and volume, respectively, whereas $B_0$ is the bulk modulus at ambient pressure and $B_0'$ is pressure derivative of the bulk modulus.
}
\begin{ruledtabular}
\begin{tabular}{ccccc}\medskip
 & $E_0$ (eV/f.u.) & $V_0$ (\r A$^3$/f.u.) & $B_0$ (GPa) & $B_0'$ \\
\textit{$\alpha$-RuBr$_3$} & & & & \\
Experiment &          & 103.7    & 19.8(6) & 6.1(7) \\\medskip
DFT        & 0.110(1) & 104.7(1) & 18.6(3)  & 6.6(2) \\
\textit{$\alpha$'-RuBr$_3$} & & & & \\
Experiment &          & 101.5    & 17.7(3) & 6.3(2) \\\medskip
DFT        & 0.143(1) & 101.5(1) & 19.3(5) & 7.8(3) \\ 
\textit{$\beta$-RuBr$_3$} & & & & \\
Experiment &     & 107.4(2) & 14.0(5) & 6.5(2) \\
DFT        & 0   & 107.6(1) & 17.7(3) & 5.7(1) \\
\end{tabular}
\end{ruledtabular}
\end{table}

All of the RuBr$_3$ polymorphs feature low bulk moduli $B_0$ of less than 20\,GPa at ambient pressure and a strong tendency to hardening upon compression, with the $B_0'$ values well exceeding the typical range of $B_0'=4-5$. This elastic behavior is characteristic of van der Waals solids. For example, the $R\bar 3$ polymorph of BiI$_3$ features $B_0=11.7(4)$\,GPa and $B_0'=8.1(3)$~\cite{schwarz2019}. 

\begin{figure*}
	\includegraphics[width=\textwidth]{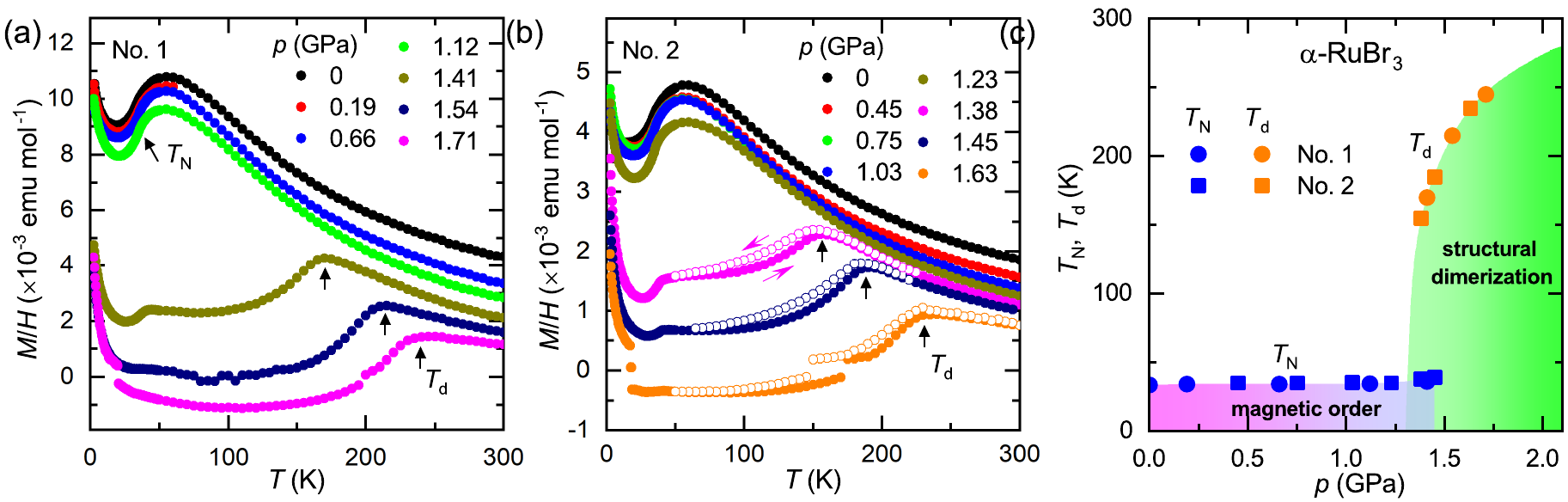}
		\vspace{-12pt} \caption{\label{fig:magnetic} Temperature-dependent dc magnetic susceptibility $M/H$ for $\alpha$-RuBr$_3$ measured at various pressures from 2\,K to 300\,K in the 1\,T magnetic field for (a) run No. 1 and (b) No. 2. The labels $T_{\rm{N}}$ and $T_{\rm{d}}$ denote the antiferromagnetic ordering transition and the structural dimerization temperature, respectively. Solid symbols were collected upon warming, while open symbols were measured upon cooling. Small discontinuities in the 1.63\,GPa and 1.71\,GPa data are due to the signal crossing zero. Such zero crossings should not occur in a paramagnetic sample and arise from the weak diamagnetic background caused by the incomplete subtraction of the signal from the pressure cell. (c) Temperature-pressure phase diagram of $\alpha$-RuBr$_3$.}
	\vspace{-12pt}
\end{figure*}

Our data suggest that $\beta$-polymorph should be the thermodynamically stable form of RuBr$_3$ at ambient pressure (Fig.~\ref{fig:energies}a). Only at lower volume does it become less stable than $\alpha'$-RuBr$_3$. The magnetic $\alpha$-polymorph is never the lowest-energy phase. Its formation becomes possible only because the $\alpha'-\beta$ transformation is kinetically hindered, so that $\alpha'$-RuBr$_3$ can be quenched and gives way to the $\alpha$-polymorph at lower pressures. It is worth noting that the $\alpha'$- and $\beta$-polymorphs show very similar bulk moduli, as seen from their almost parallel $V(P)$ curves (Fig.~\ref{fig:volume}a). The lower volume of the $\alpha'$-polymorph renders it more stable under pressure.

Transition pressures are quantified by enthalpies calculated using pressure-dependent volumes extracted from the equation of state. Fig.~\ref{fig:energies}b shows that the $\alpha-\alpha'$ transition should take place at 1.7\,GPa in a good agreement with the zero-temperature value of 1.3\,GPa determined from the magnetization measurements (Sec.~\ref{sec:magnetic}). The XRD data show the transition at the higher pressure of 2.5\,GPa at room temperature. This shift of the transition pressure with temperature should be caused by the phonon contribution to the free energy, which was not included in our DFT calculation. 

The $\beta-\alpha'$ transition is expected around 5.5\,GPa, but it could not be observed in our room-temperature XRD experiment because this transformation involves a major structural reorganization from chains into layers of the RuBr$_6$ octahedra. We argue that such a transformation must be kinetically hindered and requires elevated temperatures to be completed.

\subsection{Magnetic properties}
\label{sec:magnetic}

Fig.~\ref{fig:magnetic} shows the dc magnetic susceptibility $M/H$ as a function of temperature measured under various pressures in two separate runs. In both runs, the susceptibility displays a maximum around 55\,K at low pressures, followed by a kink at around $T_N\simeq 35$\,K that signals the formation of long-range antiferromagnetic order~\cite{imai2022}. Both features are rather robust against small pressures. By tracking the position of the kink at $T_N$ using the peak in $dM/dT$~\cite{supplement}, we find that the ordering temperature of $\alpha$-RuBr$_3$ weakly increases under pressure with the slope of $dT_N/dP\simeq 1.8$\,K/GPa. This increase is comparable to the changes observed in other Kitaev magnets, such as $\alpha$-Li$_2$IrO$_3$ with $dT_N/dP\simeq 1.5$\,K/GPa~\cite{shen2022}. Remarkably, $\alpha$-RuCl$_3$ shows an opposite trend, the reduction in $T_N$ upon compression with $dT_N/dP\simeq -13.6$\,K/GPa~\cite{wolf2022}.

Above 1.3\,GPa, the broad susceptibility maximum disappears, while a more asymmetric feature appears at $T_d$ and rapidly shifts toward higher temperatures with increasing pressure. Concurrently, the susceptibility decreases and even becomes diamagnetic above 1.6\,GPa because of the enhanced background signal as the gasket is compressed. The feature at $T_d$ is accompanied by a temperature hysteresis and indicates a first-order phase transition that can be assigned to the $\alpha-\alpha'$ structural phase transition observed by XRD. First-order nature of this transition is further corroborated by an intermediate region with the phase coexistence. Both $T_N$ and $T_d$ can be observed between 1.3 and 1.5\,GPa, as shown in Fig.~\ref{fig:magnetic}b.

Our magnetization data confirm paramagnetic nature of $\alpha'$-RuBr$_3$, as expected from its dimerized structure. The critical pressure of the $\alpha-\alpha'$ transition is strongly temperature-dependent, similar to other Kitaev magnets~\cite{bastien2018,veiga2019,shen2021,shen2022}. This temperature dependence is rooted in the different phonon spectra of the two polymorphs. Phonon contribution additionally stabilizes the magnetic, nondimerized phase at elevated temperatures~\cite{shen2022}, thus shifting the $\alpha-\alpha'$ transition toward higher pressures. Indeed, at room temperature this transition occurs at 2.5\,GPa, as seen from our XRD data.


\subsection{Electronic structure}

\begin{figure}
\includegraphics{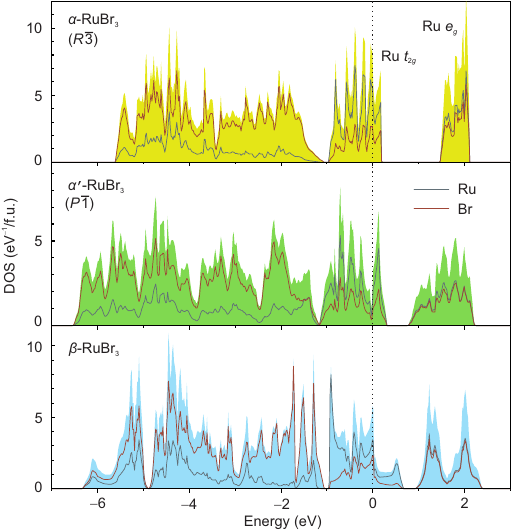}
\caption{\label{fig:dos}
Density of states (DOS) calculated without spin polarization on the DFT+SO level for the RuBr$_3$ polymorphs. The Fermi level is at zero energy. Ambient-pressure crystal structures of $\alpha$- and $\beta$-polymorphs are used. In the case of $\alpha'$-RuBr$_3$, we used the optimized crystal structure at 1.9\,GPa, right above the transition.
}
\end{figure}

Fig.~\ref{fig:dos} compares electronic structures of three RuBr$_3$ polymorphs calculated on the DFT+SO level. All of them show the broad valence band predominantly formed by the Br $4p$ states below $-1$\,eV followed by two distinct complexes of the Ru $t_{2g}$ and $e_g$ bands, with the former lying near the Fermi level in agreement with the $4d^5$ electronic configuration of Ru$^{3+}$. 

In contrast to the iridates~\cite{winter2017}, Ru$^{3+}$ compounds do not show a clear splitting of the $t_{2g}$ bands into the $j_{\rm eff}=\frac32$ and $j_{\rm eff}=\frac12$ states expected in the relativistic case, because the spin-orbit coupling constant of $\lambda\simeq 0.15$\,eV for Ru$^{3+}$ is relatively small compared to the band width. Nevertheless, the difference between the $\alpha$- and $\alpha'$- polymorphs is clearly visible in the vicinity of the Fermi level. The DOS of $\alpha'$-RuBr$_3$ features a pseudogap at the Fermi level formed as a result of transforming Ru $t_{2g}$ states into molecular orbitals driven by the short Ru--Ru bonds (dimers). Therefore, $\alpha$-RuBr$_3$ would become insulating only upon adding electronic correlations that split the $t_{2g}$ bands, whereas the band gap opening in $\alpha'$-RuBr$_3$ is almost completed by the formation of the Ru--Ru bonds. We can thus classify $\alpha$-RuBr$_3$ as Mott insulator, whereas $\alpha'$-RuBr$_3$ should be proximate to a band insulator. 

These assignments are corroborated by the DFT+$U$+SO calculations with $U_d=2$\,eV and $J_H=0.26$\,eV~\cite{gretarsson2024} that produce the magnetic insulating state for $\alpha$-RuBr$_3$ with the band gap of 0.6\,eV and 
Ru magnetic moment of 1.07\,$\mu_B$ comprising almost equal spin and orbital contributions of about 0.55\,$\mu_B$ each. Experimentally, insulating behavior of $\alpha$-RuBr$_3$ has been reported at ambient pressure~\cite{sato2024}. 


Another consequence of the Ru--Ru dimer formation is the broadening of the $e_g$ bands as a result of the enhanced Ru--Br interactions. On the level of ligand-field theory for a RuBr$_6$ octahedron, these $e_g$ bands can be thought as the Ru--Br antibonding states, whereas the respective bonding states occur at the bottom of the valence band, around $-5$\,eV. Such bonding states extend to lower energies when the metal-metal bonds are formed. All these features are similar across the $\alpha'$- and $\beta$-polymorphs despite their different crystal structures. 

\subsection{Magnetic interactions}

\begin{figure}
\includegraphics{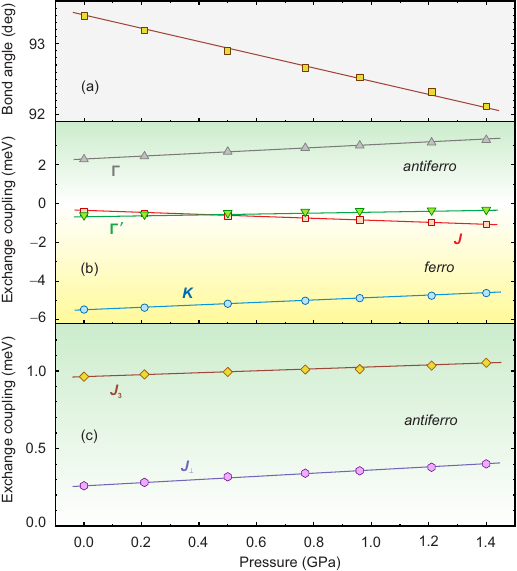}
\caption{\label{fig:exchange}
Pressure dependence of the Ru--Br--Ru bond angle (a) and exchange couplings (b,c) for $\alpha$-RuBr$_3$. The lines are guides for the eye. 
}
\end{figure}

To analyze pressure evolution of magnetism within the nondimerized honeycomb phase, we model $\alpha$-RuBr$_3$ using the extended Kitaev $J-K-\Gamma-\Gamma'$ Hamiltonian for nearest-neighbor couplings in the honeycomb plane~\cite{rousochatzakis2024},
\begin{align}
 \mathcal H=\sum_{\langle ij\rangle}\left(J\mathbf S_i\mathbf S_j+KS_i^{\gamma}S_j^{\gamma}+\Gamma(S_i^{\alpha}S_j^{\beta}+S_i^{\beta}S_j^{\alpha})+\right. \notag \\
 +\left.\Gamma'(S_i^{\alpha}S_j^{\gamma}+S_i^{\gamma}S_j^{\alpha}+S_i^{\beta}S_j^{\gamma}+S_i^{\gamma}S_j^{\beta})\right),
\end{align}
which is augmented by interactions beyond nearest neighbors ($J_3$) as well as interlayer couplings ($J_{\perp}$). The interaction parameters are obtained from the superexchange model of Refs.~\cite{rau2014,winter2016} using the on-site Coulomb repulsion $U_d=2$\,eV, Hund's coupling $J_H=0.26$\,eV, and spin-orbit coupling constant $\lambda=0.15$\,eV~\cite{gretarsson2024}. We find that $\alpha$-RuBr$_3$ is dominated by the $K<0$ and $\Gamma>0$ terms, whereas $J$ and $\Gamma'$ are both negative and less than 1\,meV in magnitude (Fig.~\ref{fig:exchange}b). The main sub-leading term is the third-neighbor in-plane interaction $J_3$ followed by the shortest interlayer coupling $J_{\perp}$, which is perpendicular to the honeycomb layers. At ambient pressure, $\alpha$-RuBr$_3$ features $K\simeq -5$\,meV, $\Gamma\simeq 2.5$\,meV, and $J_3\simeq 1.0$\,meV, which is comparable to the results of the earlier \textit{ab initio} study~\cite{kaib2022} and remarkably similar to the parameter regime established for $\alpha$-RuCl$_3$~\cite{winter2017b,wang2017,laurell2020,suzuki2021}. This parameter regime places $\alpha$-RuBr$_3$ into the region of collinear zigzag order in agreement with the experiment~\cite{imai2022}.

Pressure reduces the Ru--Br--Ru bond angle and enhances $\Gamma$, while reducing the absolute value of $K$. Concurrently, $J$ slightly increases in magnitude (and remains negative), whereas $\Gamma'$ slightly decreases (Fig.~\ref{fig:exchange}b). Such changes are consistent with the expected evolution of the nearest-neighbor couplings on reducing the bond angle~\cite{winter2016} and mainly arise from the increased diagonal hoppings, i.e., the hoppings $t_1$ and $t_3$ between $d$-orbitals of the same symmetry, similar to the Kitaev iridates~\cite{kim2016}. These trends also mirror pressure evolution of the exchange couplings in $\alpha$-RuCl$_3$, although much larger changes were proposed in that case~\cite{wolf2022}.

\begin{figure}
\includegraphics{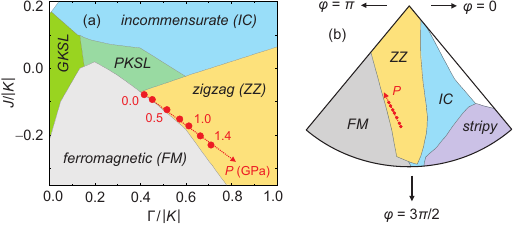}
\caption{\label{fig:diagram}
Pressure evolution of RuBr$_3$ with respect to competing phases of the extended Kitaev model. (a) Quantum phase diagram of the $J-K-\Gamma$ model ($\Gamma'=0$)~\cite{wang2019}. GKSL and PKSL are the generic and proximate Kitaev spin liquids, respectively. (b) Classical phase diagram of the $J-K-\Gamma$ model with $\Gamma'/A=-0.05$~\cite{rau2014b}. The diagram is drawn in polar coordinates: $J/A=\sin\theta\cos\varphi$, $K/A=\sin\theta\sin\varphi$, $\Gamma/A=\cos\theta$ where $0<\theta<\pi/2$ gauges the distance from the center of the circle and $A=\sqrt{J^2+K^2+\Gamma^2}$ is the overall energy scale.
}
\end{figure}

The dissimilar trends in $\Gamma$ and $K$ suggest that the overall energy scale gauged by $\sqrt{J^2+K^2+\Gamma^2+\Gamma'^2}$ remains almost constant, 5.94\,meV at 0\,GPa vs. 5.77\,meV at 1.4\,GPa, so it can't be the reason for the increase in $T_N$ under pressure. Using our exchange parameters, we place $\alpha$-RuBr$_3$ onto phase diagrams of the extended Kitaev model, the quantum phase diagram obtained for $\Gamma'=0$~\cite{wang2019} and classical phase diagram with the small $\Gamma'$~\cite{rau2014b}. In both cases, RuBr$_3$ straddles the boundary between the ferromagnetic and zigzag states while moving away from the Kitaev point because $|K|/\Gamma$ decreases (Fig.~\ref{fig:diagram}). 

The increasing $T_N$ indicates an additional stabilization of the zigzag state under pressure, which could hardly be explained by the nearest-neighbor couplings alone. The zigzag order in $\alpha$-RuBr$_3$ and other Kitaev magnets is further stabilized by $J_3$~\cite{winter2016}. This coupling is enhanced under pressure and increases by 7\% at 1.4\,GPa. The interlayer coupling $J_{\perp}$ increases by 40\% in the same pressure window. Both trends could serve to explain the experimentally observed 7\% increase in $T_N$ between 0 and 1.4\,GPa. It remains unclear why $T_N$ increases in $\alpha$-RuBr$_3$ but decreases in $\alpha$-RuCl$_3$ under pressure~\cite{wolf2022}. In both compounds, the nearest-neighbor exchange couplings evolve in a very similar way. Therefore, the terms beyond nearest neighbors, $J_3$ and $J_{\perp}$, will most likely determine the $T_N$ value, but the evolution of these terms in $\alpha$-RuCl$_3$ has not been reported and remains an interesting topic for future investigation. 

\section{Discussion and Summary}

Our data show that the Kitaev candidate $\alpha$-RuBr$_3$ is energetically less favorable than the $\beta$-polymorh and thus metastable at ambient pressure. However, the $\alpha\rightarrow\beta$ transformation is kinetically hindered and does not occur at room temperature. This allows the quenching of the honeycomb polymorph. Its synthesis from $\beta$-RuBr$_3$ requires not only high pressures but also elevated temperatures. The compression of $\beta$-RuBr$_3$ at room temperature does not suffice, as our experiments have shown, even though the transition is thermodynamically favored above 5.5\,GPa. 

The higher stability of $\beta$-RuBr$_3$ at ambient pressure is likely rooted in the large size of bromine and the more sparse nature of the chain-like structure, as opposed to the layered one. Likewise, the more compact nature of the layered structure renders it thermodynamically stable at elevated pressures. Synthesis of $\alpha$-RuBr$_3$ involves the $\beta\rightarrow\alpha'$ transformation, followed by the conversion of the $\alpha'$-polymorph into $\alpha$-RuBr$_3$ upon release of pressure. Only at this last step does the material become magnetic, whereas the first step of the $\alpha$-RuBr$_3$ synthesis, the transformation between the chain and layered polymorphs, occurs between the two structures that both contain Ru dimers. The higher stability of these polymorphs can be traced back to the enhanced Ru--Br bonding (Fig.~\ref{fig:dos}). The magnetic $\alpha$-polymorph exists in a narrow pressure window up to 1.3\,GPa at 0\,K and up to 2.5\,GPa at 295\,K. This pressure window is nevertheless much broader than in $\alpha$-RuCl$_3$ where dimerization sets in already at 0.1\,GPa at low temperatures~\cite{wolf2022}. 

The nonmagnetic dimerized phases of RuBr$_3$ and RuCl$_3$ are structurally similar. The bromide is more robust against magnetism collapse than the chloride in agreement with the larger size of bromine and the larger unit-cell volume at ambient pressure. \textit{Ab initio} calculations suggest very similar pressure evolution of nearest-neighbor exchange couplings in both compounds. It comes then as a surprise that the trends in $T_N$ are different: whereas magnetic ordering is stabilized by pressure in $\alpha$-RuBr$_3$, it is strongly suppressed in $\alpha$-RuCl$_3$~\cite{wolf2022}. The increasing $T_N$ of the bromide can be rationalized by the enhancement of the third-neighbor in-plane as well as interplane couplings. A detailed analysis of these couplings in the chloride is clearly warranted.

From the magnetism perspective, pressure tunes $\alpha$-RuBr$_3$ away from the Kitaev limit because it reduces the Ru--Br--Ru bond angles and enhances $\Gamma$ while reducing $|K|$. These changes do not visibly affect the position of $\alpha$-RuBr$_3$ with respect to the boundary between the ferromagnetic and collinear zigzag states, so pressure evolution of $T_N$ is more likely to be affected by the third-neighbor in-plane and interplane couplings, both increasing under pressure. Bringing $\alpha$-RuBr$_3$ closer to the Kitaev limit would require an expansion of the structure via negative pressure. Partial iodine substitution~\cite{sato2024,ni2024} and strain tuning~\cite{kaib2021} may be useful in this context.

In summary, we have studied pressure evolution of the different RuBr$_3$ polymorphs and revealed magnetism collapse of the honeycomb $\alpha$-polymorph, the structural sibling of the renowned $\alpha$-RuCl$_3$. Our data suggest that the pressure-induced transformation from the chain structure into the honeycomb structure takes place between the two paramagnetic phases that both contain the Ru--Ru dimers. These dimers disappear upon release of pressure, giving way to the magnetic $\alpha$-polymorph, which is metastable. The pressure window of this magnetic $\alpha$-RuBr$_3$ is somewhat larger than in the chloride, but it does not exceed that of the Kitaev iridates. Hydrostatic pressure tunes $\alpha$-RuBr$_3$ away from the Kitaev limit, so that expansion rather than contraction of the structure would be necessary in order to enhance Kitaev interactions in this material.

Experimental and computational data associated with this manuscript are available from Refs.~\cite{esrf,zenodo}.

\acknowledgments
We acknowledge ESRF for providing beamtime for this project and Anton Jesche for his technical and conceptual support. AT thanks Ioannis Rousochatzakis for fruitful discussions. This work was funded by the Deutsche Forschungsgemeinschaft (DFG, German Research Foundation) -- TRR 360 -- 492547816 (subproject B1). B.S. and P.K.M. acknowledge the financial support of Alexander von Humboldt Foundation

%


\clearpage\newpage
\begin{widetext}
\begin{center}
\large\textbf{\textit{Supplemental Material}\smallskip \\ Magnetic versus nonmagnetic polymorphs of RuBr$_3$ under pressure}
\end{center}
\end{widetext}

\renewcommand{\thesection}{S\arabic{section}}
\renewcommand{\thefigure}{S\arabic{figure}}
\renewcommand{\thetable}{S\arabic{table}}
\setcounter{figure}{0}
\setcounter{table}{0}
\setcounter{section}{0}

\begin{widetext}

\section{Additional structural data}

\begin{figure*}[!b]
\includegraphics{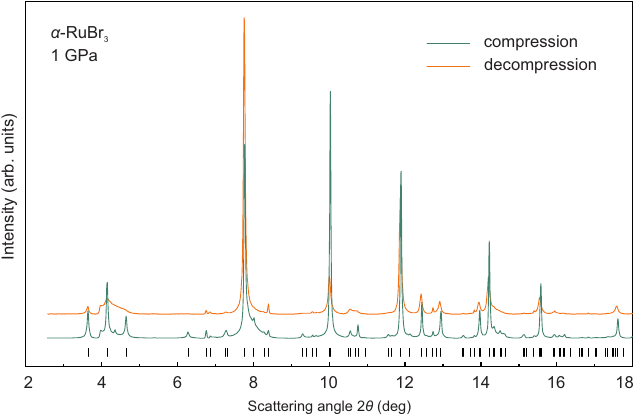}
\caption{\label{fig:patterns}
XRD patterns of $\alpha$-RuBr$_3$ collected around 1\,GPa upon compression and decompression. The patterns are offset for clarity.
}
\end{figure*}

Fig.~\ref{fig:patterns} shows the XRD patterns of $\alpha$-RuBr$_3$ collected upon compression and decompression. They demonstrate reversibility of the $\alpha-\alpha'$ transition. However, the crystallinity of the sample is notably reduced after pressure cycling.

Fig.~\ref{fig:lattice} displays pressure-dependent lattice parameters of $\beta$-RuBr$_3$. At room temperature and ambient pressure, $\beta$-polymorph has an orthorhombic ($Pmmn$) crystal structure with the ordered arrangement of the Ru--Ru dimers~\cite{hillebrecht2004,prots2023}. Above 384\,K, it reversibly transforms into a hexagonal ($P6_3/mcm$) structure where similar chains with the alternation of short and long Ru--Ru distances become disordered relative to each other~\cite{hillebrecht2004}. Low crystallinity of our $\beta$-RuBr$_3$ sample did not allow us to resolve the orthorhombic superstructure. Therefore, we analyzed the XRD data for $\beta$-RuBr$_3$ using hexagonal symmetry and determined two lattice parameters, $a$ and $c$, that reflect the interchain and intrachain distances, respectively. 

\begin{figure*}
\vspace{0.4cm}
\includegraphics{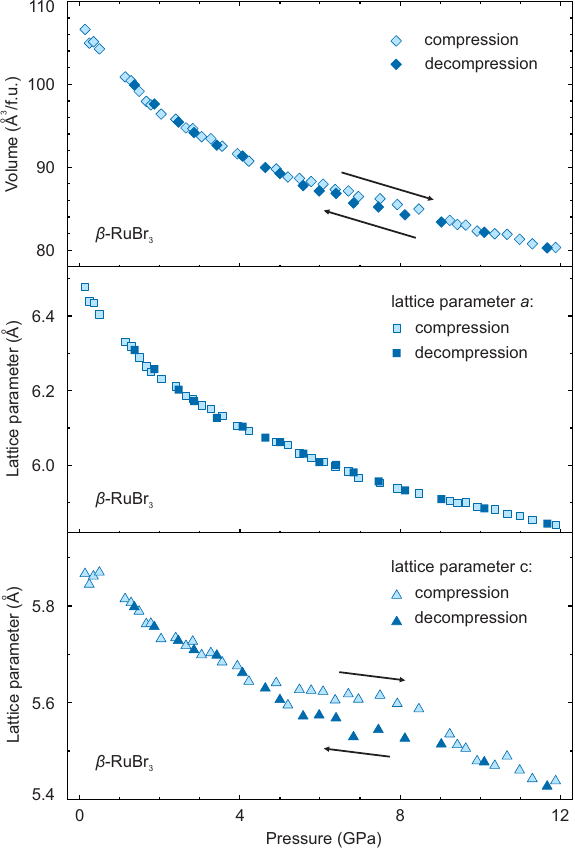}
\caption{\label{fig:lattice}
Pressure dependence of the unit-cell volume and lattice parameters for $\beta$-RuBr$_3$.
}
\end{figure*}

Pressure evolution of $\beta$-RuBr$_3$ reveals a small hysteresis between 5 and 10\,GPa indicating a possible phase transition associated with the shortening of the $c$ parameter. Details of this transition could not be resolved in the present experiment, owing to the low crystallinity of the $\beta$-RuBr$_3$ sample. However, it is worth noting that the transition happens in the pressure range where the $c$-parameter approaches 5.6\,\r A, which is about twice the value of the Ru--Ru intradimer distance at ambient pressure ($2.71-2.73$\,\r A~\cite{hillebrecht2004,prots2023}). Therefore, it is plausible that the discontinuity in the $c$ lattice parameter of $\beta$-RuBr$_3$ corresponds to a transformation within the dimerized Ru chains, indicating for example the loss of dimerization, but further experiments performed on a sample with higher crystallinity would be necessary in order to determine the Ru--Ru distances as a function of pressure and verify this conjecture.

\section{Experimental determination of $T_N$}

\begin{figure*}
\includegraphics[width=10cm]{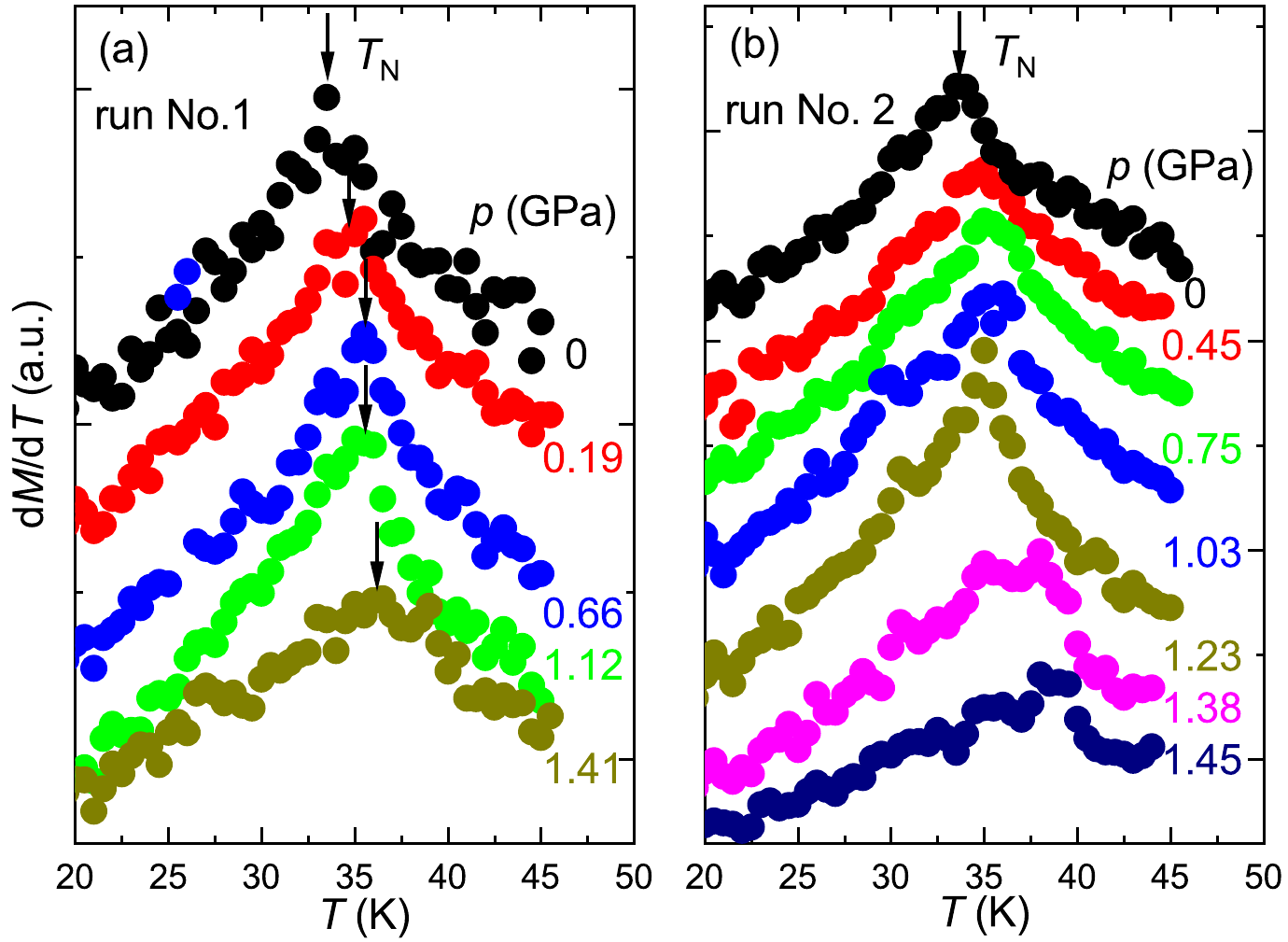}
\caption{\label{fig:TN}
Determination of $T_N$ using $dM/dT$ measured at different pressures.
}
\end{figure*}

Fig.~\ref{fig:TN} shows temperature derivative of the magnetization, $dM/dT$, to illustrate pressure evolution of $T_N$. The peak in $dM/dT$ identifies the kink in $M(T)$ that signals phase transition associated with the magnetic ordering.

\section{Pressure evolution of electron hoppings}

Fig.~\ref{fig:hoppings} displays pressure evolution of the electron hoppings extracted using Wannier projections. The hoppings are written in the conventional form of
\begin{equation*}
 \begin{array}{l@{\hspace{0.3cm}}|@{\hspace{0.5cm}}c@{\hspace{0.5cm}}c@{\hspace{0.5cm}}c}
 & d_{yz} & d_{xz} & d_{xy} \\\hline
 d_{yz} & t_1 & t_2 & t_4 \\
 d_{xz} & t_2 & t_1 & t_4 \\
 d_{xy} & t_4 & t_4 & t_3 \\
\end{array}
\end{equation*}
The individual Ru--Ru bonds in $\alpha$-RuBr$_3$ are not subject to any symmetry, so the two diagonal hoppings labeled with $t_1$ are not strictly equal. The same holds true for the four off-diagonal hoppings identified as $t_4$. However, deviations from the aforementioned form of the hopping tensor are minor, less that 5\%, thus justifying the description of $\alpha$-RuBr$_3$ in terms of the $J-K-\Gamma-\Gamma'$ model. The values of $t_1$ and $t_4$ shown in Fig.~\ref{fig:hoppings} are averaged over the respective nonequivalent hoppings.

\begin{figure*}
\includegraphics{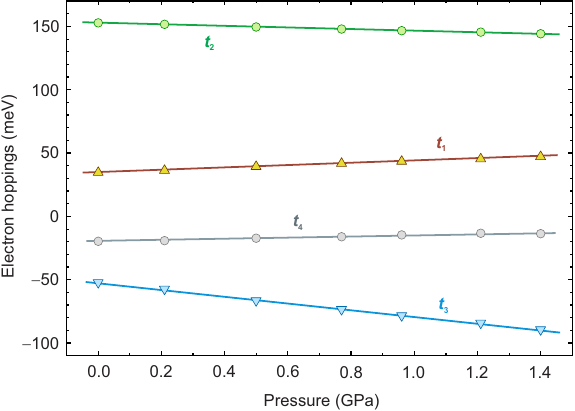}
\caption{\label{fig:hoppings}
Pressure dependence of electron hoppings. The lines are guides for the eye.
}
\end{figure*}

\end{widetext}

\end{document}